# Educational Short Videos: Bibliometric Trends, Thematic Structure, and Operationalisation

Yidi Zhang[1], Jindi Wang[2], Pedro Bem-Haja[1], Zhipeng Li[2], Lei Shi[3], Thomas K. F. Chiu[4]

[1] University of Aveiro [2] Eastern Institute of Technology, Ningbo

[3] Newcastle University [4] The Chinese University of Hong Kong

## Abstract

**Background:** Educational short-video research spans disciplines, platforms and learning contexts, but the same label is applied to resources differing in function, activity, context and evaluation, complicating comparison and evidence synthesis.
**Objectives:** This study mapped the development, thematic organisation and operationalisation of educational short-video research.
**Methods:** We analysed 2,169 records indexed in Web of Science and Scopus up to 11 June 2026 using bibliometric analysis, non-negative matrix factorisation topic modelling and structured content analysis.
**Results and Conclusions:** Publication output increased sharply from the mid-2010s but remained dispersed across outlets. Among 16 first-level topics, Skill Development in Educational Contexts was the largest, forming the structural core of the field, while topic overlap was predominantly pairwise. Video-Based Health Interventions for Attitude Change and Cognitive Load and Engagement in Instructional Video Design combined positive recent growth with comparatively high citation visibility, whereas Social Media Engagement Strategies for Education emerged as a rapidly growing direction. Knowledge/Achievement and Engagement/Motivation were the most widely represented outcome domains, and experimental and synthesis designs were more common among identifiable records in health-related topics. Indexed descriptions most often foregrounded intended users, educational uses, learning content, and interactivity. A representative duration was available for 511 records, but no consistently applied numerical threshold was evident. These findings support a working definition of educational short videos as discrete multimedia messages combining words and visuals and designed or used to promote learning. Shortness should be interpreted relative to educational purpose, content unit and surrounding activity rather than duration alone.

## Lay Summary

### What is currently known about this topic?

- Educational short-video research is divided across disciplines, platforms and learning contexts.
- Existing reviews usually isolate subjects, platforms or pedagogical formats.
- The same label covers videos with fundamentally different educational functions and activities.

### What does this paper add?

- Sixteen topics reveal a structured field centred on skill development.
- Topics follow distinct growth, citation, outcome and study-design profiles.
- Shortness has no consistent numerical boundary and is defined through educational function.

### Implications for practice and/or policy

- Evidence syntheses should not pool studies solely because videos are labelled “short”.
- Researchers should report video purpose, learner activity, context, outcomes and duration together.
- Educators should design each video around a coherent learning unit, not a minute limit.

# 1. Introduction

Video has become a routine instructional resource in face-to-face, blended, and online learning environments because it can combine verbal explanation with visual representation to support learning (Mayer, 2020; Zhang et al., 2006). Its educational value, however, depends not only on the information presented but also on how that information is selected, organised, segmented, and paced, as well as how learners interact with and exercise control over the material (Choe et al., 2019; Mayer, 2021; Merkt & Bodemer, 2024). Within this broader development of video-based learning, short videos and related concise formats have attracted growing attention as flexible resources for teaching, learning, training, and knowledge communication (Brame, 2016; Guo et al., 2014). This interest reflects both instructional concerns with focused and manageable video resources and the wider expansion of short-form video across contemporary digital media environments (Conde-Caballero et al., 2024; Nguyen & Diederich, 2023; Otto, 2025; Wang et al., 2024).

Educational short videos are used across diverse settings, including language learning, management and science education, medical and health education, professional training, and platform-based learning (Hu et al., 2025; Mooring et al., 2016; Nguyen & Diederich, 2023; Wang et al., 2024; Xie et al., 2025; Zhang et al., 2023). They may function as instructor-designed explanations or demonstrations, pre-class resources in flipped learning, learner-generated artefacts, professional training or health-education resources, and publicly accessible educational content (Brame, 2016; Campbell et al., 2022; Frydenberg & Andone, 2016; Zahirović Suhonjić et al., 2019). Taken together, these applications show that educational short videos constitute a heterogeneous category rather than a uniform educational intervention.

This heterogeneity has important implications for the interpretation of research findings. Some studies have associated shorter videos or particular production formats with greater learner engagement or more favourable learning outcomes in specific contexts (Guo et al., 2014; Slemmons et al., 2018), whereas others have raised concerns about possible associations between intensive short-video use and rational thinking, cognitive functioning, learning or academic performance (Otto, 2025; Xu et al., 2023). These studies examine different objects, the former typically focus on deliberately designed instructional resources, whereas the latter may address more general patterns of platform use or exposure. Even where an identifiable educational purpose is present, instructor explanations, procedural demonstrations, learner-generated artefacts and platform-based public education videos may differ substantially in educational purpose, content structure, learner role and surrounding activity (Brame, 2016; Campbell et al., 2022; Hu et al., 2025; Nguyen & Diederich, 2023).

This heterogeneity is also reflected in definitions of educational short video and in the ways existing reviews have organised the literature. Zhu et al. (2023) noted the absence of a uniform definition of short video and argued that the format should not be defined by duration alone. At the same time, existing reviews have generally delimited their scope according to particular instructional contexts, platforms, communicative purposes or adjacent pedagogical formats. Zhang et al. (2022) reviewed short videos in foreign-language teaching and learning, Tan et al. (2022) examined the pedagogical potential of TikTok in English as a second language classrooms, and Yang et al. (2025) extended the platform-based perspective across several higher-education disciplines while remaining focused on TikTok applications, learning outcomes and implementation factors. In health communication, Zhu et al. (2023) synthesised evidence on the persuasive effects of short videos, whereas Li et al. (2024) reviewed methods for assessing the quality of health science-related short videos on TikTok. Microlearning reviews offer a broader related perspective, but their scope also includes quizzes, flashcards and brief texts and therefore does not isolate educational short video as a distinct object of analysis (Monib et al., 2025). Taken together, existing reviews organise the literature according to different inclusion logics and do not provide a field-level synthesis across disciplines, platforms and learning contexts. Educational short-video research has therefore not yet been examined as a coherent field in terms of its thematic structure, development patterns, evidence composition and definitional boundaries.

To address this gap, the present study integrates bibliometric analysis, graded non-negative matrix factorisation topic modelling and structured content analysis in a corpus of 2,169 records from Web of Science and Scopus. It maps the development and thematic organisation of the field, compares topics in terms of connectivity, temporal development, citation visibility, study-design composition and outcome coverage, and examines how educational short videos have been operationalised through reported duration and educational characteristics. By bringing these dimensions into a single analytical framework, the study provides a field-level basis for assessing cross-study comparability and for distinguishing research that shares terminology from research that examines educationally comparable resources.

The following research questions guided the study:

RQ1. How has educational short-video research developed in terms of publication output, citation visibility, outlet distribution and international collaboration?

RQ2. What major topics structure educational short-video research, and how do they differ in connectivity, temporal development, citation visibility, reported study-design composition and outcome coverage?

RQ3. How have video duration and educational characteristics been reported and operationalised in indexed descriptions of educational short-video research?

# 2. Materials and Methods

## 2.1 Analytical Corpus

### 2.1.1 Data Sources and Search Strategy

The identification, deduplication, screening and inclusion processes are summarised in a PRISMA-informed flow diagram (Figure 1). Publication records were retrieved from the Web of Science Core Collection and Scopus. Searches covered the Topic field in Web of Science and the title, abstract, and keyword fields in Scopus. These databases were selected because they provide broad coverage of research across education, technology, communication and other disciplines in which educational short-video studies may be published. The complete database-specific search strategies, including all spelling, spacing, pluralisation and hyphenation variants, are provided in Appendix I.

The searches covered records indexed up to 11 June 2026 and yielded 2,980 records from Web of Science and 3,624 from Scopus, producing an initial pool of 6,604 records before deduplication. As the 2026 search covered only the period up to 11 June, that year was treated as a partial observation in all temporal analyses.

### 2.1.2 Eligibility Criteria

For study identification, educational short video was treated as a literature-defined working category. No disciplinary restrictions were applied. Records were eligible when: (a) the publication explicitly referred to a short video or related concise video format; and (b) the focal video served an identifiable purpose involving learning, teaching, training, public education or another educational activity.

No universal duration cut-off was imposed because one aim of the study was to examine how shortness had been defined and operationalised across the literature. Records were excluded when the focal material did not involve an identifiable educational use of short video, including entertainment, marketing, general media use or platform exposure, technical video processing and non-educational communication. Education-related records were also excluded when short video or video-based microlearning was not the focal object.

Inclusion in the corpus therefore indicated a reported educational use of a short or otherwise concise video format, it did not imply equivalence among the included videos in duration, conceptual scope or pedagogical design.

### 2.1.3 Deduplication and Screening

Records retrieved from Web of Science and Scopus were merged and deduplicated using exact DOI and normalised-title matching. DOIs were standardised by removing URL prefixes and formatting variations, while titles were normalised for case, punctuation and whitespace. Candidate duplicates sharing an identical DOI or normalised title were manually verified using author names, publication year and journal or source information; only confirmed duplicates were removed. No fuzzy matching or similarity threshold was applied. This process yielded 4,384 unique records for screening.

Two reviewers independently screened the titles, abstracts and keywords of all records against the eligibility criteria. Additional bibliographic information was consulted when the educational function of the video or the focal video format could not be determined from the indexed fields alone.

The reviewers agreed on 4,008 of the 4,384 screening decisions, corresponding to 91.4% agreement and Cohen's $\kappa = .83$. The remaining 376 discrepant or borderline records were reassessed against the eligibility criteria and resolved through discussion between the two reviewers.

A total of 2,215 of the 4,384 screened records were excluded (50.5%). These comprised 76 records without abstracts available for the title–abstract–keyword analyses (3.4% of all exclusions), 1,853 records without an identifiable educational use of short videos (83.7%), and 286 education-related records in which short video or video-based microlearning was not the focal object (12.9%). The disciplinary subcategories within the second exclusion category are shown in Figure 1. The final analytical corpus comprised 2,169 records.

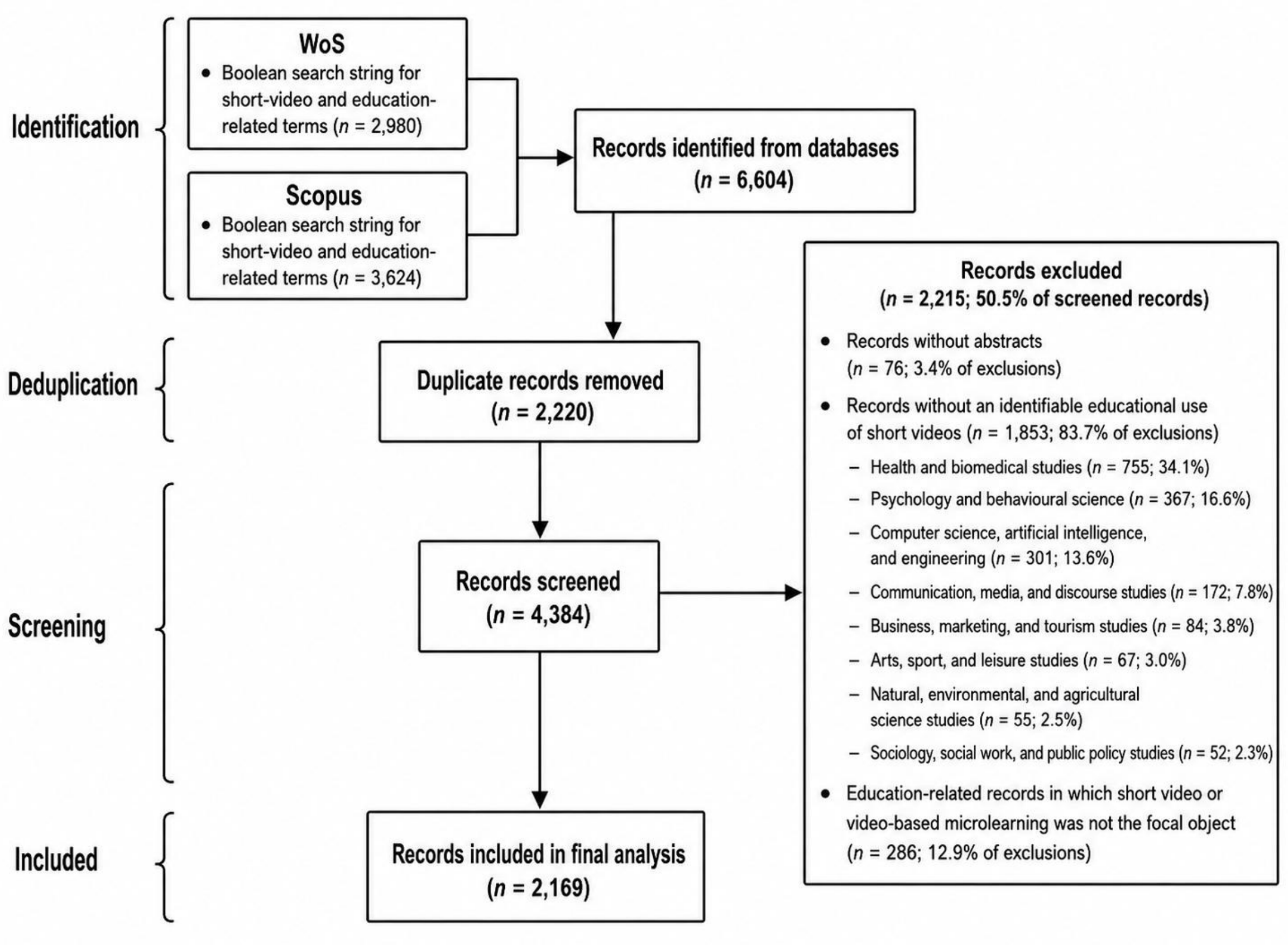

Figure 1. Flow diagram of record identification, deduplication, screening, exclusion and inclusion. Percentages for the three main exclusion categories and their disciplinary subcategories were calculated using all excluded records (n = 2,215) as the denominator.

## 2.2 Bibliometric and Affiliation-Based Co-Authorship Analysis

To address RQ1, annual publication output, citation visibility, outlet distribution and affiliation-based international collaboration were analysed. Annual publication counts were calculated for the full corpus. Citation visibility was assessed separately using Scopus and Web of Science citation counts. For each database, a record-level annualised citation value was calculated by dividing the citation count by publication age, defined as (2026 − publication year + 1). This denominator provided a consistent year-based approximation of citation exposure across the corpus. As a robustness check, the analysis was repeated after excluding records published between 2024 and 2026, which had the shortest citation windows. These measures were interpreted as indicators of database-specific citation visibility rather than research quality or educational effectiveness. For Bradford-style zoning, outlets were ranked by publication count and divided into three zones, each accounting for approximately one third of the corpus. Outlets tied on publication count were ordered alphabetically to ensure reproducibility.

International collaboration was analysed using the country or region information reported in author affiliations. Full counting was applied, such that each country or region represented in a publication received one publication count, irrespective of the number of authors from that location. Countries or regions were treated as network nodes, and an edge was created when two locations appeared in the same publication. Edge weights represented the number of publications shared by each pair. The network was described using the number of countries or regions, distinct links, network density, connected components, isolates, collaboration partners and the most frequent country or region pairs.

## 2.3 Topic Modelling

### 2.3.1 Corpus Construction and Text Preprocessing

To address RQ2, the thematic structure of the final corpus (N = 2,169) was modelled from the title, abstract, and combined-keywords fields of each record. The three fields were concatenated at the record level before preprocessing. Text was converted to lowercase and normalised for Unicode accents. URLs, punctuation, and repeated whitespace were removed, together with standard English stop words and a corpus-specific stop-word list. The processed text was represented by term frequency–inverse document frequency (TF–IDF)-weighted unigrams and bigrams. Additional computational details, including TF–IDF construction, topic-number diagnostics and selection, topic-label generation and auditing, and topic-prevalence calculations, are provided in Appendix II. The use of titles, abstracts and keywords provided a consistent textual representation across the complete corpus. Accordingly, the resulting topics represent themes foregrounded in indexed publication records rather than exhaustive representations of the full-text content of individual studies.

### 2.3.2 NMF Estimation and Graded Topic Membership

Non-negative matrix factorisation (NMF) was fitted to the TF–IDF matrix:

$$X \approx WH, \qquad W \geq 0, \quad H \geq 0,$$

where W contains record–topic weights and H contains topic–term weights. Models were estimated using NNDSVDa initialisation, coordinate-descent optimisation, Frobenius loss, a random seed of 42, and a maximum of 1,000 iterations. Each row of W was divided by its row sum to obtain relative topic-membership weights:

$$\theta_{dk} = \frac{W_{dk}}{\sum_{j=1}^{K} W_{dj}}.$$

These weights sum to one within each record. They were used as graded, non-exclusive memberships and were not interpreted as posterior probabilities.

### 2.3.3 Selection of the Number of Topics

Candidate solutions were fitted for K = 6, …, 30. Four diagnostics were compared: topic coherence, topic diversity, topic-term stability, and multi-topic coverage. The eligible candidate with the highest composite heuristic score was retained. This rule selected K = 16 (Figure 2): coherence = 0.292, diversity = 0.865, stability = 0.710, multi-topic coverage = 0.775, distinctiveness = 0.764, and maximum pairwise redundancy = 0.253. These values are descriptive diagnostics rather than inferential statistics.

The 16 NMF components were retained without post-estimation merging. K = 16 was retained in 90.5% of 243 plausible heuristic specifications, while analyses of adjacent solutions (K = 14–18) showed high subsample stability and preserved the principal prevalence and temporal findings for RQ2. Full diagnostic procedures and sensitivity results are reported in Appendix III (Figures III.1–III.3).

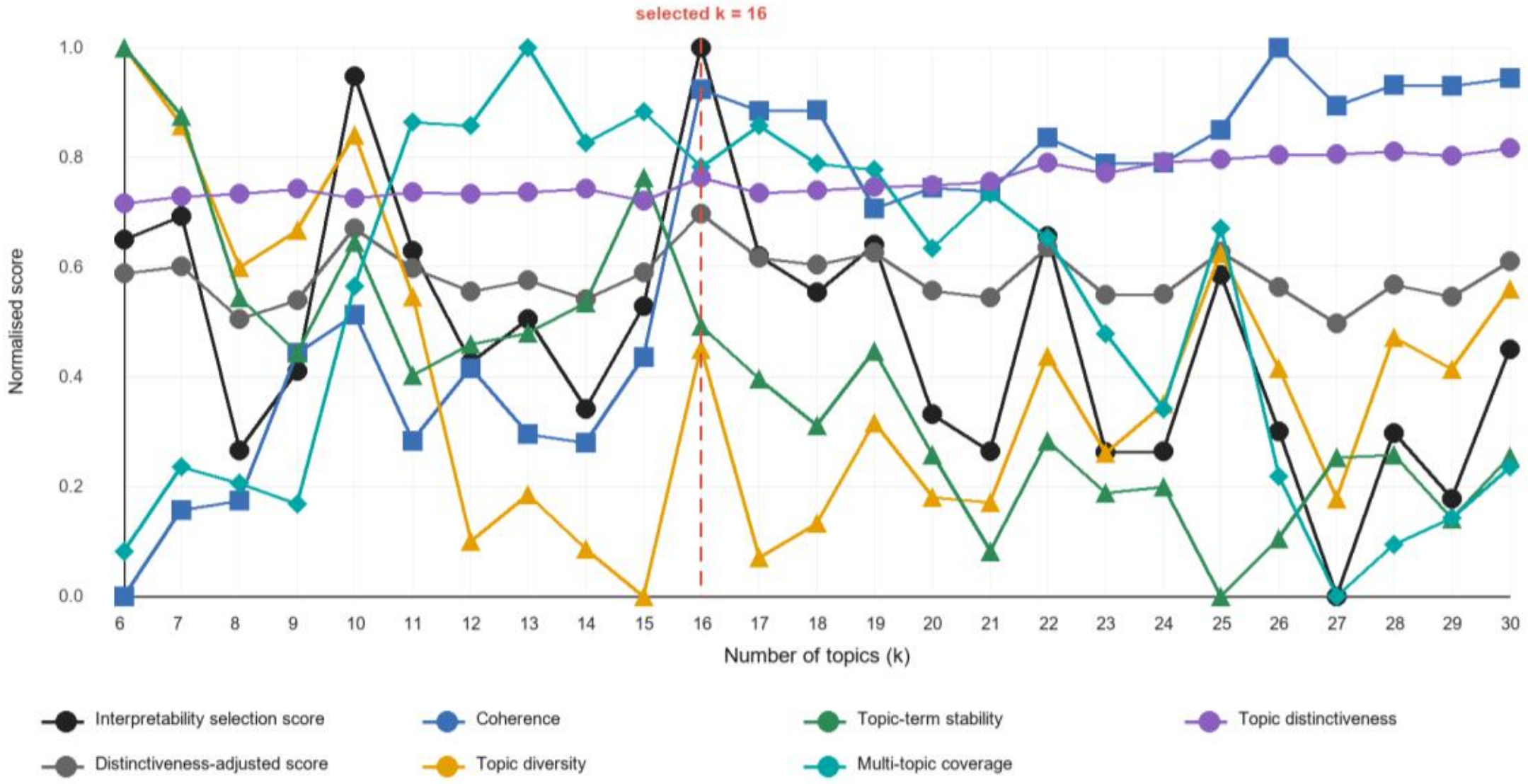

Figure 2. Diagnostics for candidate NMF solutions (K = 6–30). The vertical dashed line marks the retained K = 16 solution. Each diagnostic was min–max normalised across the candidate solutions; the composite line is a heuristic selection score.

### 2.3.4 Topic Interpretation and Record Membership

Topic formation and topic naming were treated as separate analytical stages. The number and composition of the topics were fixed by the NMF model-selection procedure described above. The retained K = 16 solution was not subsequently merged, split, or re-estimated during interpretation. Label development could therefore change only the verbal description of a component and not its topic-term weights, record-topic weights, or membership structure.

The substantive scope of each fixed component was examined using its 12 highest-weighted terms and the titles of the eight records with the largest component weights. The same evidence was submitted independently to gpt-4o-mini through the OpenAI API and deepseek-chat through the DeepSeek API. In a second OpenAI API call, the 16 provisional labels were compared jointly to identify semantic overlap and improve the verbal distinction between neighbouring topics. This contrastive step addressed label redundancy only and did not alter the NMF components.

Two researchers independently evaluated each provisional topic label against the 12 highest-weighted terms, the eight representative record titles, the scope statement, and neighbouring topic labels. Differences in interpretation or wording were resolved through discussion, and labels were revised where necessary to improve specificity and distinguish neighbouring topics. This study affected only the descriptive labels and did not alter the NMF solution, topic weights, record memberships, thresholds, or prevalence estimates.

The APIs were used as labelling aids rather than as topic estimators, simulated annotators, or substitutes for researcher judgement. Final responsibility for topic interpretation and manuscript wording remained with the researchers. For descriptive reporting, a record was linked to topic k when $\theta_{dk} \geq 0.15$. When more than three topics met this threshold, only the three largest weights were retained.

### 2.3.5 Topic Evolution and Co-occurrence

Ordinary least-squares slopes were fitted for the full period ending in 2025 and, separately, for 2021–2025. The slopes were treated as descriptive summaries of the direction and rate of change rather than as causal or forecasting estimates. Records from 2026 were shown as partial-year observations and excluded from both slope estimates.

For the main co-occurrence analysis, a topic was counted as present when its weight was at least 0.20. Two topics co-occurred when both met this criterion in the same record. Pairwise and three-topic counts were inclusive: a record with three qualifying topics contributed to all three embedded pairs and to the corresponding triplet. Sensitivity analyses repeated the procedure at thresholds of 0.10, 0.15, 0.20, and 0.25. The 0.20 threshold was retained for the main display because it removed many links based on small residual weights while retaining the multi-topic structure. Threshold-specific coverage is reported with the sensitivity results. The ten most frequent pairs and triplets were displayed.

Topic-level network connectivity was represented by node strength, calculated as the sum of the weighted co-occurrence links connecting each topic to all other topics. Larger values therefore indicated that a topic participated more frequently in thresholded co-occurrences across the corpus. Because topic memberships were derived from lexical patterns in titles, author keywords and abstracts, the network represents overlap in indexed textual content rather than theoretical integration or disciplinary cross-pollination. Node strength is therefore interpreted only as a descriptive measure of weighted record-level topic co-occurrence.

#### 2.3.6 Second-Level Topic Modelling within Overlapping Parent Topics

To examine within-topic heterogeneity, separate NMF models were fitted within the overlapping corpora associated with each first-level topic. Records with a parent-topic weight of at least 0.15 entered the corresponding parent corpus and could therefore be included in more than one second-level analysis. The number of second-level topics was selected using corpus-size restrictions and a heuristic combining topic distinctiveness, term diversity, distributional balance and minimum subtopic share. Second-level topics were used descriptively and did not alter the first-level topic solution, record memberships or quantitative comparisons. Full model-selection procedures, diagnostic criteria and labelling procedures are reported in Appendix II.

## 2.4 Study-Design and Outcome-Domain Classification

To examine how reported study designs and outcome- and evaluation-related domains varied across topics, we constructed an exploratory topic–outcome reporting landscape. Classification was conducted using a deterministic rule-based procedure applied to the titles, author keywords and abstracts of all 2,169 records. The procedure was analytically independent of the LLM-assisted topic-labelling process, and each assigned category was accompanied by the expression that triggered the classification, providing a record-level audit trail.

Research design was represented using an exploratory ordinal indicator based on the highest design category explicitly reported in the indexed fields. The predefined ordering distinguished systematic syntheses, randomised experiments, controlled experiments or quasi-experiments, pre–post or evaluative interventions, longitudinal or analytical observational studies, mixed-methods or cross-sectional studies, qualitative or content analyses, and descriptive or developmental reports. Indicator values ranged from 0.5 for descriptive or developmental reports to 4.0 for systematic syntheses and randomised experiments. No value was assigned when study design could not be identified. Study design was identifiable for 1,053 records (48.5%).

The indicator was used as a pragmatic summary of the relative representation of reported study-design categories. It was not interpreted as a validated hierarchy of evidence or as an assessment of methodological quality, risk of bias, certainty of evidence or intervention effectiveness.

Outcome- and evaluation-related reporting was classified using seven non-exclusive, study-specific dictionaries: Knowledge/Achievement, Engagement/Motivation, Cognition/Attention, Skills/Procedural Performance, Reflection/Professional Development, Assessment/Feedback, and Content Quality/Accuracy. These categories were developed as pragmatic reporting domains rather than as an exhaustive taxonomy of learning outcomes. Content Quality/Accuracy represents a resource-level evaluation domain, while the remaining categories capture learner outcomes, learning processes, or assessment-related constructs. Records could contribute to more than one outcome domain. At least one outcome domain was identified for 1,788 records (82.4%).

For each topic–domain cell, weighted record volume was calculated as the sum of the records' graded NMF topic-membership weights. The study-design profile was summarised as the topic-weighted mean indicator among records with an identifiable design. In Figure 10, bubble size represents weighted record volume, while colour represents the corresponding mean indicator value. Thus, the two visual encodings describe weighted record volume and reported design composition separately.

Because the classifications relied on indexed metadata rather than full-text coding, they identify study designs and outcome domains explicitly foregrounded in titles, abstracts and keywords. Design details, secondary outcomes and experimental procedures reported only in the full text may therefore have been missed. The resulting matrix is interpreted as an exploratory description of study-design and outcome-related information explicitly reported in indexed metadata. The absence of a metadata-based match should not be interpreted as evidence that the corresponding design feature or outcome was absent from the full study.

## 2.5 Extraction of Video Duration

To address RQ3, representative video duration was obtained from accessible bibliographic information and open-access full texts. Only explicit, quantifiable statements referring to the duration of the video itself were included. Durations referring to total learning time, class or session length, intervention period, platform use, assessment, or follow-up were excluded. Representative duration was extractable for 511 records, representing 23.6% of the full corpus. These records constituted the complete set meeting the duration-extraction criteria and were used to describe reported video durations.

Duration values were standardised to seconds and reported in minutes. When a study included multiple videos, one representative value was derived from the reported mean or median, the average of individual durations, the total duration divided by the number of videos, or the midpoint of a reported range. Records without a justifiable numerical estimate were excluded. The distribution was summarised using the median, interquartile range, mean, standard deviation, and observed range. Because duration was right-skewed, the median and interquartile range were treated as the primary statistics.

## 2.6 Extraction of Reported Educational Characteristics

Reported educational characteristics were extracted from the title, abstract, and keyword fields of all 2,169 records. These indexed fields support consistent record-level comparison but do not provide complete descriptions of the underlying videos or learning activities. The analysis therefore identified characteristics foregrounded in record descriptions rather than definitive properties of the videos.

The coding framework used the Learning Resource Metadata Initiative and corresponding Schema.org educational-resource properties as a conceptual scaffold (DCMI LRMI Task Group, 2022). Seven LRMI-derived or compatible dimensions were retained: educationalUse, teaches, assesses, educationalLevel, typicalAgeRange, educationalRole, and interactivityType. A corpus-specific dimension, platformContext, was added to capture features particularly relevant to short-video research. The complete coding dictionary and matching rules are provided in Appendix IV.

Extraction used a deterministic rule-based procedure combining exact-phrase and anchor-constrained contextual matching. Exact-phrase rules captured specific expressions such as “video quiz”, “flipped classroom video”, and “YouTube Shorts”. Contextual rules counted broader terms such as engagement, knowledge, practice, students, or classroom only when they occurred in the same sentence or local text window as a short-video-related anchor. These reduced matches generated by general educational vocabulary unrelated to the focal video.

For the internal validation of the educational-characteristics coding dictionary, 200 records were randomly selected and independently coded by two researchers. Inter-rater agreement was high (Cohen’s $\kappa = .904$), and disagreements were resolved by consensus to establish the reference standard. Against this standard, the dictionary achieved 86.4% accuracy.

This validation applied only to the educational-characteristics dictionary and should not be interpreted as validation of the separate study-design or outcome-domain classification procedures.

Each category was counted once per publication record regardless of repeated matches, whereas multiple categories within the same dimension were retained when explicitly reported. Supporting text and raw matches were preserved for auditing. The resulting frequencies are interpreted as record-level reporting patterns, not as exhaustive classifications of video design or educational practice.

# 3. Results

## 3.1 Publication Trends, Outlet Distribution, and Affiliation-Based International Co-Authorship

Annual publication output was sporadic before 2000, increased gradually during the 2000s and accelerated markedly from the mid-2010s. More than 100 records were published annually from 2015 onwards, with output reaching 168 records in 2023, 195 in 2024 and 214 in 2025. A further 111 records were indexed in 2026 by the data-collection cut-off of 11 June (Figure 3A). Because the 2026 data represented only part of the year, this value was not interpreted as evidence of a decline.

Age-normalised citation rates varied substantially across publication years in both Scopus and Web of Science (Figure 3B–C). Median rates were generally low, while each distribution contained a small number of highly cited records. The highest annualised citation rates were approximately 35.7 citations per year in Scopus and 35.3 in Web of Science. After records published between 2024 and 2026 were excluded, the overall median annualised citation rate changed only slightly in Scopus, from 0.18 to 0.21 citations per year, and remained at 0 in Web of Science. The maximum rates were unchanged. The concentration of citation visibility in a small number of records was therefore not attributable solely to the most recent publication cohorts.

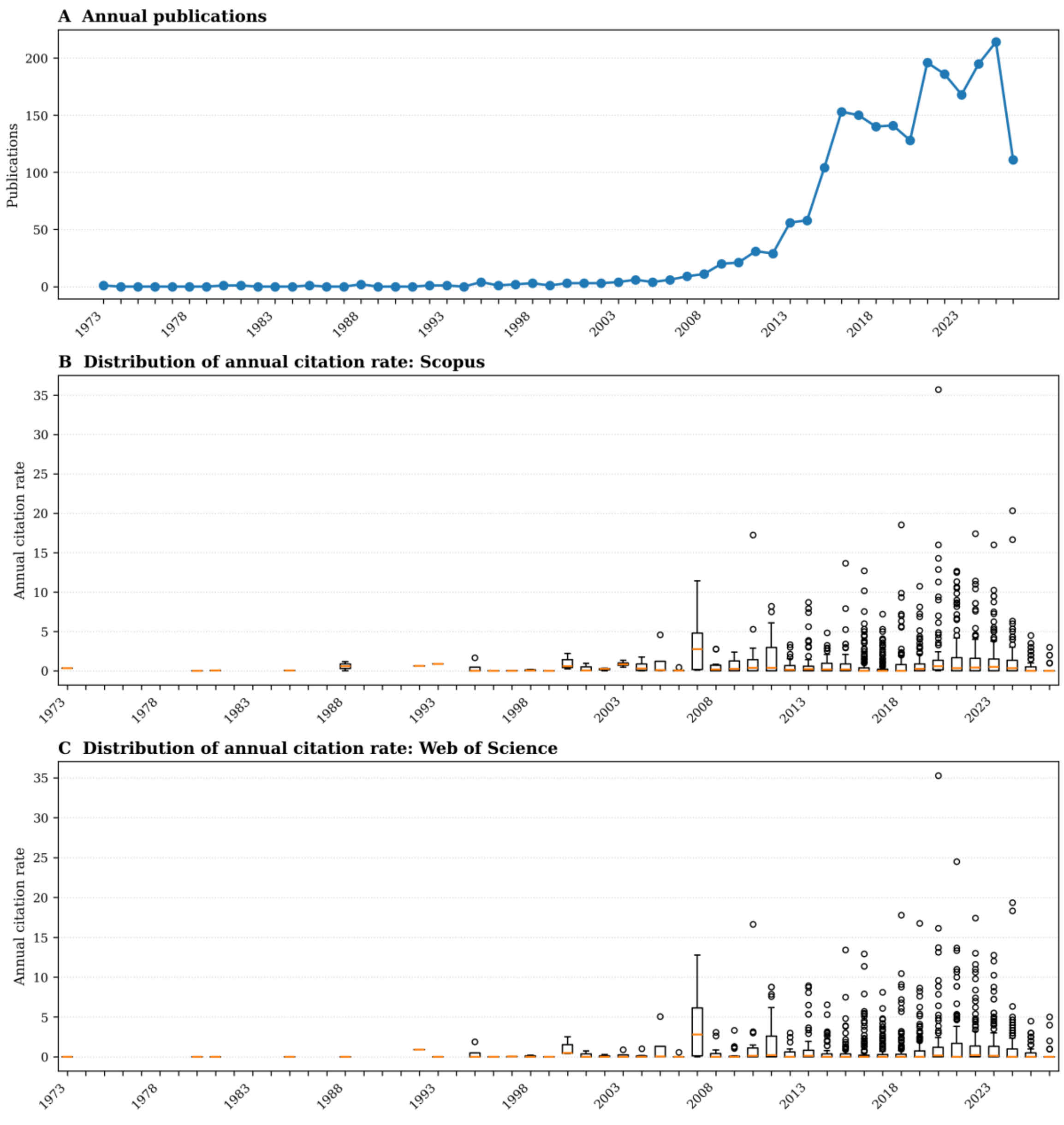

Figure 3. Annual publication output and age-normalised citation rates.

The 2,169 records were distributed across 1,483 journals and conference proceedings, indicating substantial dispersion across publication outlets. The most frequent outlet was *ASEE Annual Conference and Exposition, Conference Proceedings* with 42 records, followed by *ACM International Conference Proceeding Series* with 30 and *International Journal of Emerging Technologies in Learning* with 14. Most outlets contributed only one record, with 1,181 single-record outlets representing 79.6% of all publication venues.

Citation visibility was considerably more concentrated. *Computers & Education* received the largest number of citations in both Scopus and Web of Science, followed by *Computers in Human Behavior*, *Resuscitation*, *Language Learning & Technology*, and *Patient Education and Counseling*. Bradford-style zoning divided the corpus into three approximately equal groups by publication volume, Zone 1 contained 169 outlets and 722 records, Zone 2 contained 591 outlets and 724 records, and Zone 3 contained 723 outlets and 723 records. Despite contributing similar numbers of records, Zones 1 and 2 together accounted for 98.4% of Scopus citations and 95.7% of Web of Science citations. Zone 3 accounted for one third of the records but only 1.6% of Scopus citations and 4.3% of Web of Science citations. These results indicate that educational short-video research was widely dispersed across publication outlets, whereas citation visibility remained concentrated in a relatively small core of venues.

Affiliation metadata identified 91 countries or regions. China appeared in the largest number of records (n = 576), followed by the United States (n = 563), the United Kingdom (n = 120), Spain (n = 111) and Australia (n = 80). The international co-authorship network contained 304 distinct links and had a density of 0.0742 (Figure 4). The largest connected component included 72 countries or regions, while the full network contained 19 connected components and 17 isolates.

International participation was geographically broad, but most collaboration links were infrequent. Of the 304 links between countries or regions, 223 (73.4%) appeared in only one publication. The United States had the largest number of international partners (n = 42), followed by the United Kingdom (n = 28), Italy and Belgium (n = 25 each), and Australia (n = 24). The most frequent collaboration pairs were China–United States (n = 13), United Kingdom–United States (n = 11), Canada–United States (n = 10), China–United Kingdom (n = 8), Australia–United Kingdom (n = 7), and Australia–United States (n = 7).

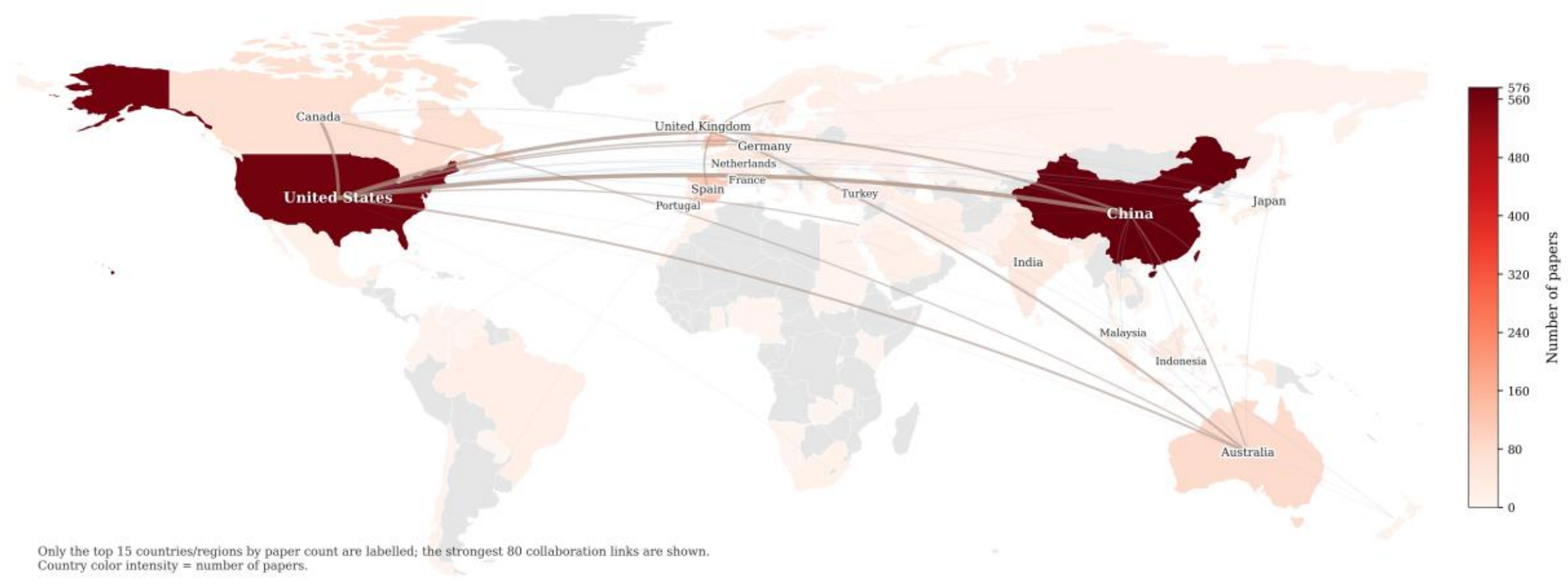

Figure 4. Affiliation-based country/region co-authorship network.

Overall, educational short-video research expanded rapidly but remained widely dispersed across publication outlets. Citation visibility was concentrated in a comparatively small group of outlets and records. International participation was extensive, although sustained collaboration was limited, as most country or region links occurred in only one publication.

## 3.2 Research Topics

The retained solution comprised 16 first-level topics. Figure 5 presents their relative corpus shares and corresponding second-level topic structure. For the descriptive shares reported in this figure, each record was assigned to the first-level topic with its highest membership weight. These dominant-topic assignments describe the overall composition of the corpus and differ from the overlapping, threshold-based memberships used in the subsequent co-occurrence analysis.

No single first-level topic accounted for more than 15% of the corpus. Skill Development in Educational Contexts was the largest topic, representing 14.4% of records. It was followed by Video-Based Health Interventions for Attitude Change (10.1%), Video-Based Patient Education in Surgical Care (9.3%), Cognitive Load and Engagement in Instructional Video Design (8.9%), Micro-Learning Strategies for Course Design (8.7%), and Innovative Approaches to Engineering Education (7.3%).

Other topics included Mobile Learning Applications for Education (6.0%), Social Media Engagement Strategies for Education (5.8%), Microlecture-Supported Flipped Classroom Pedagogy (5.4%), Enhancing Oral Skills in Foreign Language Learning (5.0%), and Design and Impact of MOOCs in Higher Education (4.6%). Micro-Learning Innovations in College English and Video Training for Bystander CPR Skills each represented 3.2% of records, followed by Short-Video Innovations in Pandemic-Era Education (3.0%), Microlearning in Higher Vocational Education (2.7%), and Innovations in Ideological Political Education (2.4%). The second-level topics and their proportions within the overlapping parent-topic corpora are also shown in Figure 5.

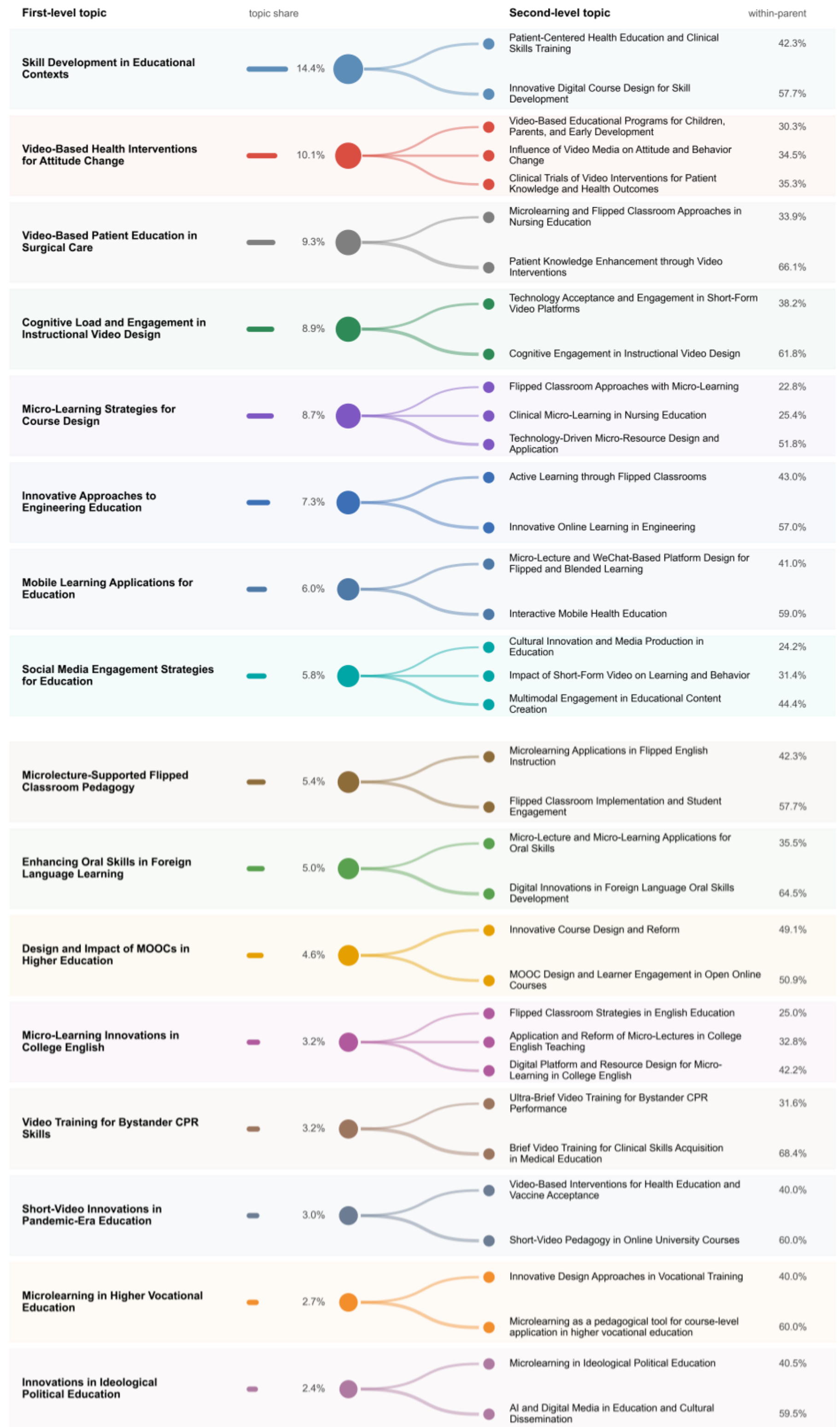

Figure 5. Two-level topic and subtopic structure of the indexed corpus, with overlapping parent-topic memberships.

Figure 6 examines thematic overlap using graded topic memberships rather than the dominant-topic assignments reported in Figure 5. A record could therefore contribute to more than one first-level topic when multiple topic weights met the specified threshold. Under this approach, Skill Development in Educational Contexts had the largest membership count (n = 599) and the highest node strength (504). The next-highest centrality values were observed for Video-Based Patient Education in Surgical Care (339), Video-Based Health Interventions for Attitude Change (304), Cognitive Load and Engagement in Instructional Video Design (281), and Micro-Learning Strategies for Course Design (252). These values indicate the frequency and strength of topic overlap in the indexed metadata and do not demonstrate theoretical integration or knowledge transfer between disciplinary areas.

The strongest pairwise overlap linked Video-Based Health Interventions for Attitude Change with Video-Based Patient Education in Surgical Care (n = 110). Other prominent links connected Video-Based Health Interventions for Attitude Change with Skill Development in Educational Contexts (n = 84), and Innovative Approaches to Engineering Education with Skill Development in Educational Contexts (n = 80). Higher-order overlap was considerably less frequent. The most common three-topic combination involved Video-Based Health Interventions for Attitude Change, Skill Development in Educational Contexts, and Video-Based Patient Education in Surgical Care (n = 10).

The sensitivity analysis showed that the number of records with multi-topic membership declined from 1,976 (91.1%) at a threshold of 0.10 to 743 (34.3%) at a threshold of 0.25. As the threshold increased, retained overlaps became increasingly concentrated in two-topic combinations.

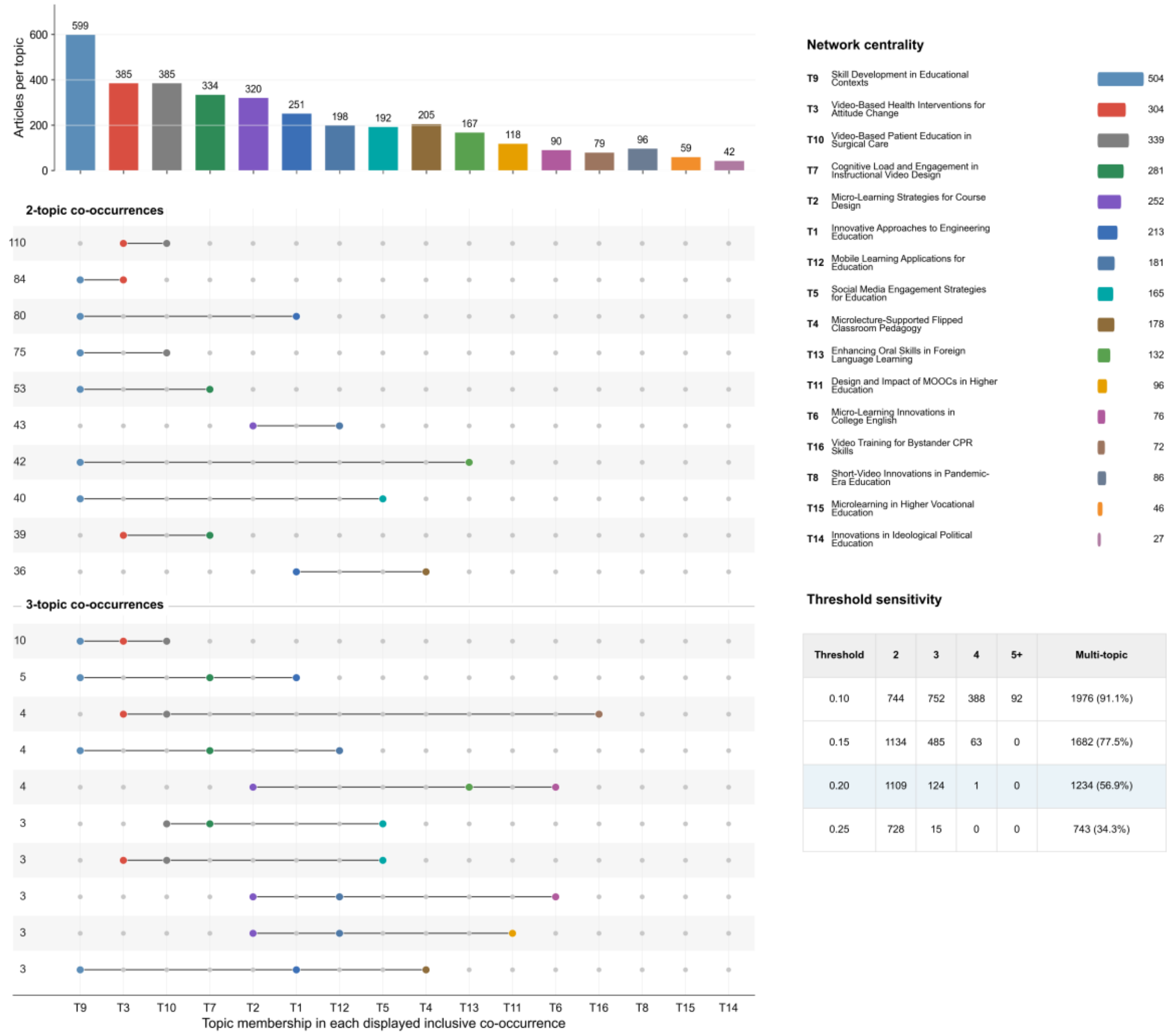

| Threshold | 2 | 3 | 4 | 5+ | Multi-topic |
|---|---|---|---|---|---|
| 0.10 | 744 | 752 | 388 | 92 | 1976 (91.1%) |
| 0.15 | 1134 | 485 | 63 | 0 | 1682 (77.5%) |
| 0.20 | 1109 | 124 | 1 | 0 | 1234 (56.9%) |
| 0.25 | 728 | 15 | 0 | 0 | 743 (34.3%) |

Figure 6. Co-occurrence structure and network centrality of first-level topics. Links represent record-level topic co-occurrences derived from titles, author keywords and abstracts; node strength represents weighted co-occurrence connectivity rather than theoretical or disciplinary integration.

Figure 7 shows changes in the relative publication shares of the 16 first-level topics. The overall slopes summarise long-term change, while the 2021–2025 slopes indicate recent publication momentum. These slopes are descriptive linear summaries of observed change, not predictive or causal estimates. Because the 2026 data covered only the period up to 11 June, they were displayed as partial-year observations and were not included in the slope estimates.

Cognitive Load and Engagement in Instructional Video Design showed the strongest recent increase, with an overall slope of +0.20 and a 2021–2025 slope of +2.67. Social Media Engagement Strategies for Education also showed substantial recent growth (+0.06; +1.80). Positive recent slopes were additionally observed for Skill Development in Educational Contexts (−0.08; +0.77), Enhancing Oral Skills in Foreign Language Learning (−0.36; +0.71), and Video-Based Health Interventions for Attitude Change (−0.03; +0.47).

Short-Video Innovations in Pandemic-Era Education showed the largest recent decline despite a positive long-term slope (+0.09; −1.78), followed by Micro-Learning Strategies for Course Design (+0.22; −1.31). Negative recent slopes were also observed for Innovative Approaches to Engineering Education (−0.11; −0.76), Design and Impact of MOOCs in Higher Education (+0.12; −0.60), Mobile Learning Applications for Education (+0.08; −0.47), Microlearning in Higher Vocational Education (+0.06; −0.45), Microlecture-Supported Flipped Classroom Pedagogy (+0.12; −0.42), Video-Based Patient Education in Surgical Care (−0.51; −0.38), and Micro-Learning Innovations in College English (+0.07; −0.32). Video Training for Bystander CPR Skills remained comparatively stable (+0.01; −0.03).

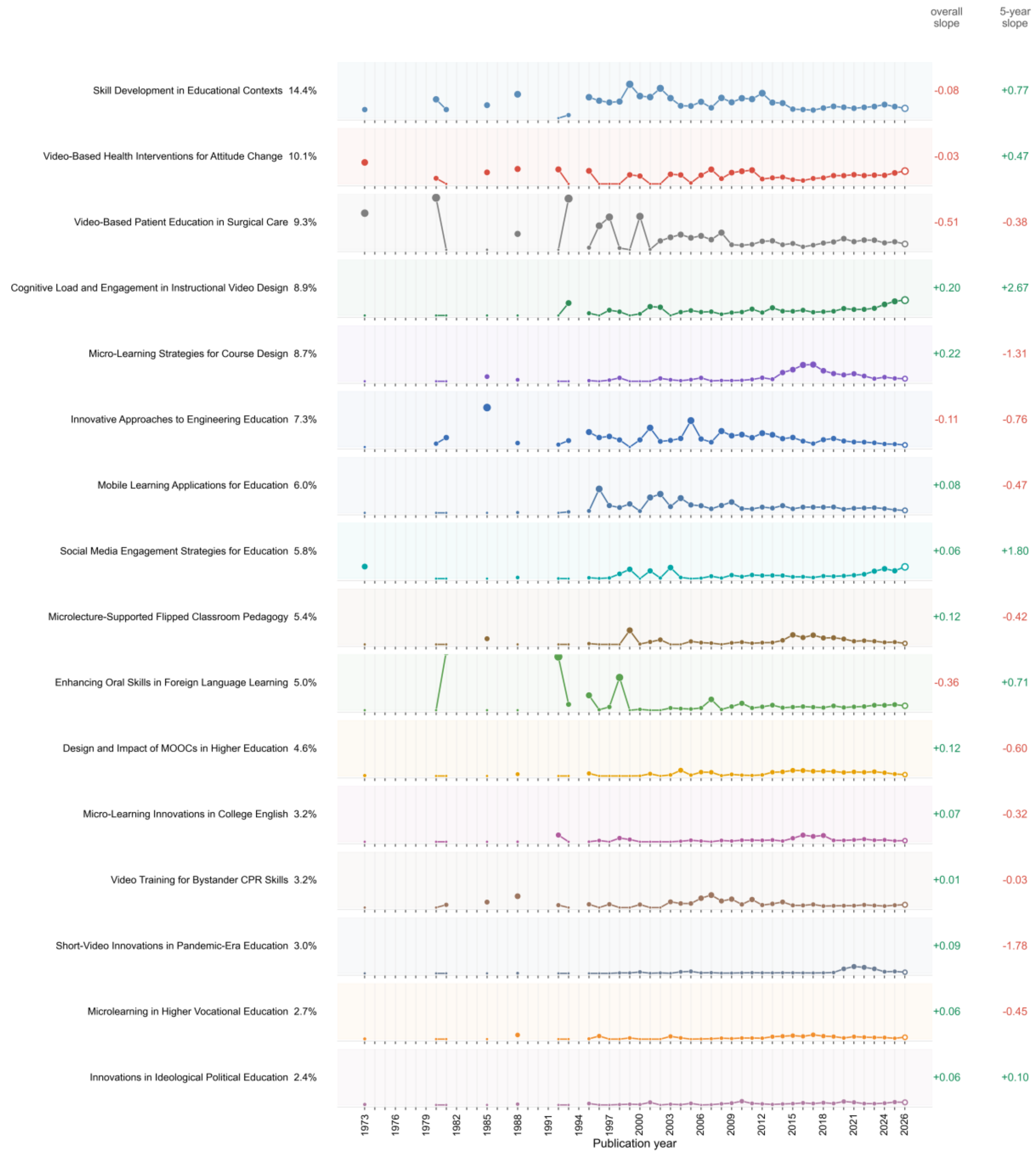

Figure 7. Long-term and recent development of first-level topics.

Figure 8 compares each topic's share of the corpus with its weighted citation shares in Scopus and Web of Science. Citation tilt was calculated as citation share minus corpus share and is reported in percentage points. Positive values indicate citation visibility above the topic's relative publication volume, whereas negative values indicate lower citation visibility relative to that volume.

Video-Based Health Interventions for Attitude Change showed the largest positive citation tilt (+8.3), followed by Cognitive Load and Engagement in Instructional Video Design (+4.5), Video Training for Bystander CPR Skills (+4.1), and Video-Based Patient Education in Surgical Care (+2.3). Enhancing Oral Skills in Foreign Language Learning (+1.6) and Design and Impact of MOOCs in Higher Education (+1.0) also received citation shares above their corresponding corpus shares.

Micro-Learning Strategies for Course Design showed the largest negative citation tilt (−6.5), followed by Skill Development in Educational Contexts (−4.0), Mobile Learning Applications for Education (−2.1), Micro-Learning Innovations in College English (−2.0), and Social Media Engagement Strategies for Education (−1.9). Short-Video Innovations in Pandemic-Era Education (−0.3) and Microlecture-Supported Flipped Classroom Pedagogy (−0.4) were closer to proportional balance between publication volume and citation visibility.

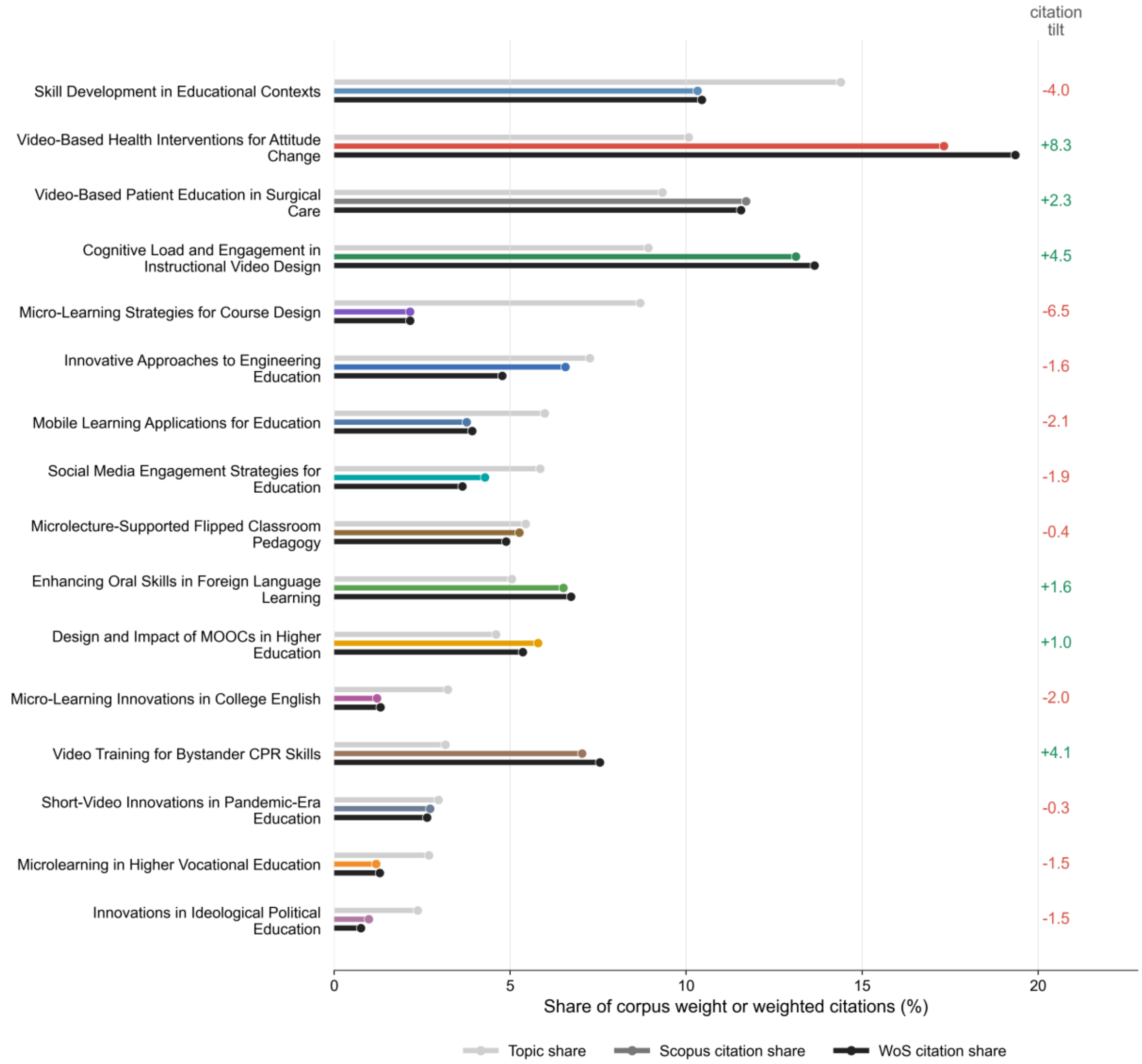

Figure 8. Topic-level citation visibility relative to corpus share.

Figure 9 integrates recent publication momentum, citation visibility, corpus share and network centrality. The horizontal axis represents the 2021–2025 slope, the vertical axis represents the mean weighted citation share across Scopus and Web of Science, bubble size represents topic share, and bubble-outline width represents network centrality.

Skill Development in Educational Contexts had the largest corpus share and the highest network centrality. Although its recent growth and citation visibility were not the highest, it had the highest weighted co-occurrence connectivity in the metadata-derived topic network. Video-Based Health Interventions for Attitude Change and Cognitive Load and Engagement in Instructional Video Design combined positive recent slopes with comparatively high citation visibility. The former had the strongest citation position, while the latter showed the greatest recent publication momentum. Enhancing Oral Skills in Foreign Language Learning also combined a positive recent slope with above-average citation visibility, although it had a smaller corpus share and lower network centrality.

Video-Based Patient Education in Surgical Care remained comparatively citation-visible and frequently co-occurred with other topics in the metadata-derived network but showed negative recent growth. Micro-Learning Strategies for Course Design accounted for a relatively large share of the corpus but showed lower recent momentum and citation visibility. Social Media Engagement Strategies for Education showed strong recent publication growth, but its citation visibility remained low relative to its publication share, indicating that the increase in publication volume had not yet been accompanied by a comparable citation share. Innovations in Ideological Political Education also showed positive recent growth, but its smaller corpus share and lower network centrality indicated less frequent overlap with other topics in the indexed metadata.

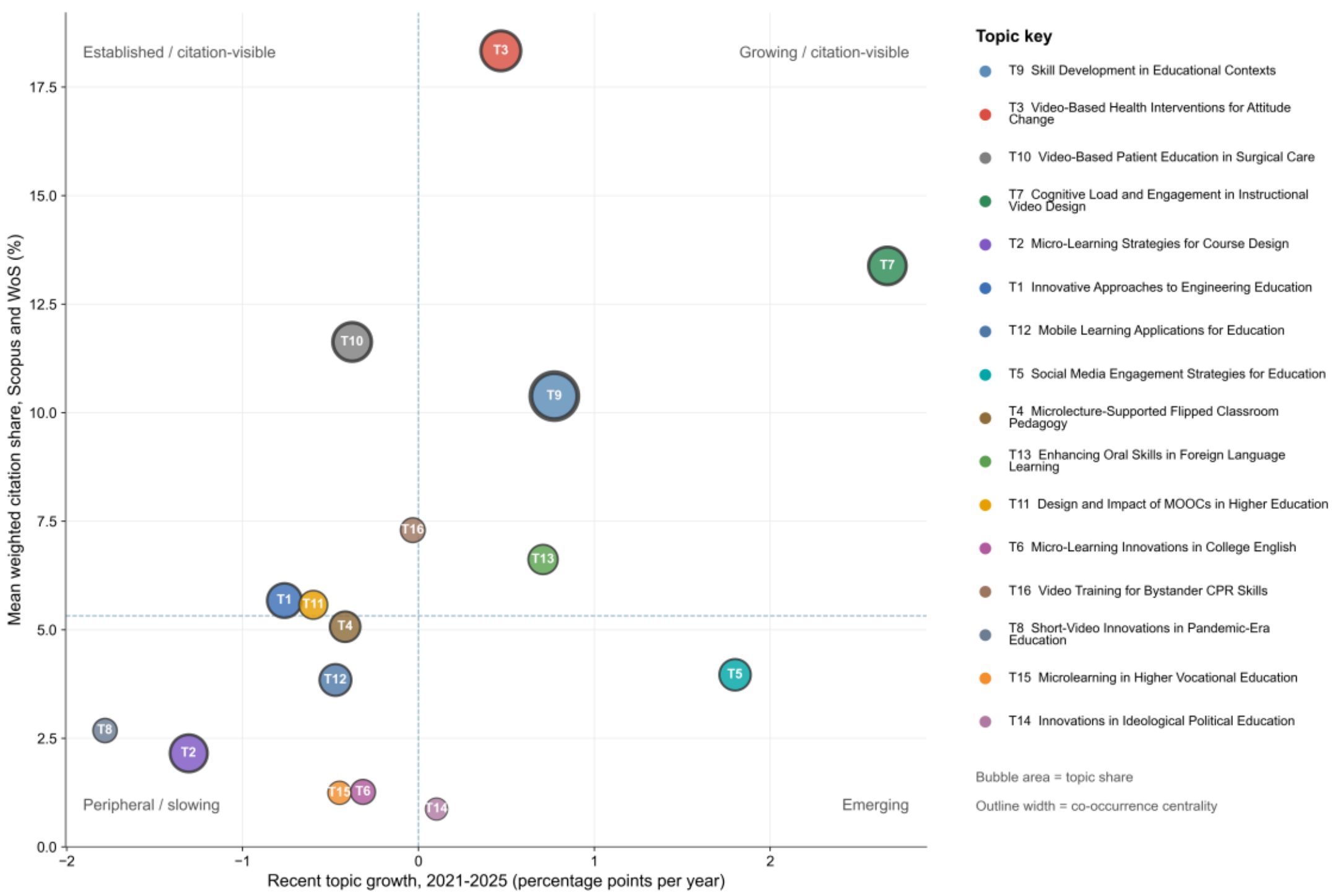

Figure 9. Integrated topic positions based on growth, citation visibility, corpus share, and network centrality.

Figure 10 shows the distribution of seven outcome- and evaluation-related reporting domains across the 16 first-level topics. Bubble size indicates topic-weighted record volume, while colour represents the topic-weighted mean of the exploratory study-design indicator, calculated only for the 1,053 records (48.5%) with an identifiable study design.

Knowledge/Achievement and Engagement/Motivation were the most widely represented outcome domains, followed by Skills/Procedural Performance and Assessment/Feedback. Reflection/Professional Development was the least represented domain across most topics. Skill Development in Educational Contexts had the broadest outcome profile, particularly in Knowledge/Achievement, Engagement/Motivation, Skills/Procedural Performance, and Assessment/Feedback. Among records assigned to this topic for which study design could be identified, descriptive and non-randomised designs were most frequently represented.

Video-Based Health Interventions for Attitude Change and Video-Based Patient Education in Surgical Care had relatively large weighted record volumes and greater representation of randomised and synthesis-level designs among records with identifiable designs. Cognitive Load and Engagement in Instructional Video Design was particularly prominent in Cognition/Attention. Video Training for Bystander CPR Skills contained fewer records but showed comparatively substantial weighted record volumes in Skills/Procedural Performance, Cognition/Attention, and Content Quality/Accuracy. Most remaining topics had smaller or more selective outcome profiles and, within their identifiable subsets, were characterised mainly by descriptive or non-randomised designs.

The reporting landscape therefore differed across both domain coverage and reported study-design composition. Weighted record volume was concentrated in Knowledge/Achievement, Engagement/Motivation, Skills/Procedural Performance, and Assessment/Feedback. Randomised and synthesis-level designs were more frequently represented among identifiable records associated with health-related topics. These patterns describe information foregrounded in indexed metadata and should not be interpreted as assessments of methodological quality, certainty of evidence, or intervention effectiveness.

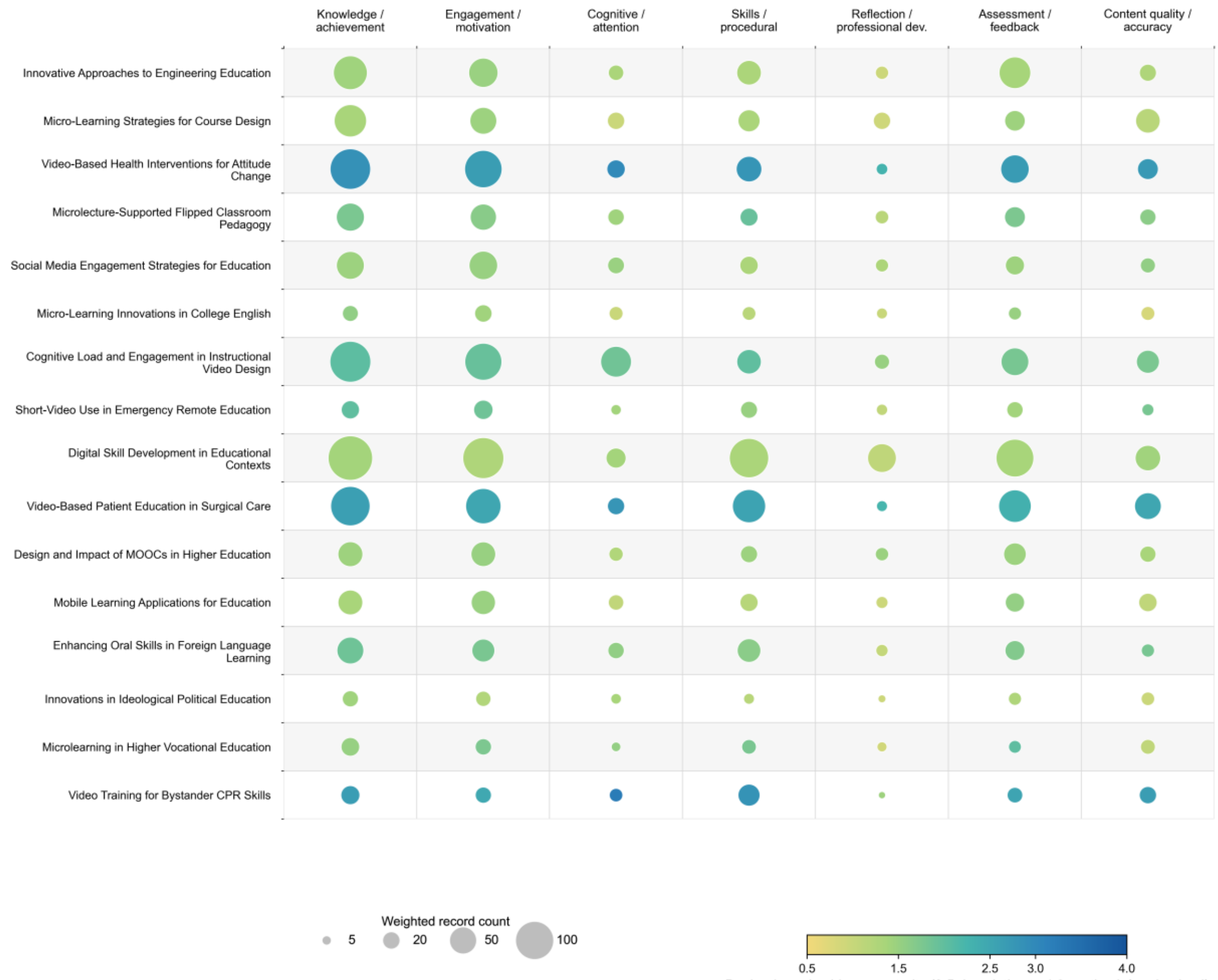

Figure 10. Reported outcome and evaluation domains and exploratory study-design indicators across first-level topics.

Taken together, the topic analyses revealed a field with a clear structural core but differentiated patterns of development, connectivity, citation visibility and evidence composition. Skill Development in Educational Contexts was the largest and most highly connected topic. Video-Based Health Interventions for Attitude Change and Cognitive Load and Engagement in Instructional Video Design combined positive recent publication momentum with comparatively high citation visibility, while Social Media Engagement Strategies for Education showed substantial recent growth but lower citation visibility relative to its publication share. Several topics concerning microlearning, mobile learning and patient education showed weaker recent momentum. Across the field, the most robust thematic overlaps were predominantly pairwise, while evidence was concentrated in knowledge and achievement, engagement and motivation, procedural skills, and assessment and feedback.

## 3.3 Defining and Operationalising Educational Short Video

To address how educational short videos were operationalised, we examined reported video duration and reported educational characteristics. The former indicates how shortness was quantified, while the latter shows how the educational functions, users, activities and contexts of the focal videos were represented in the reviewed literature.

A representative video duration was available for 511 records, corresponding to 23.6% of the full corpus (Figure 11). Durations were right-skewed, with a median of 5.00 minutes (IQR 2.32–10.00), a mean of 6.21 minutes, and an observed range of 0.08–30.00 minutes. This broad distribution did not reveal a clear and consistently applied duration cut-off within the available subset.

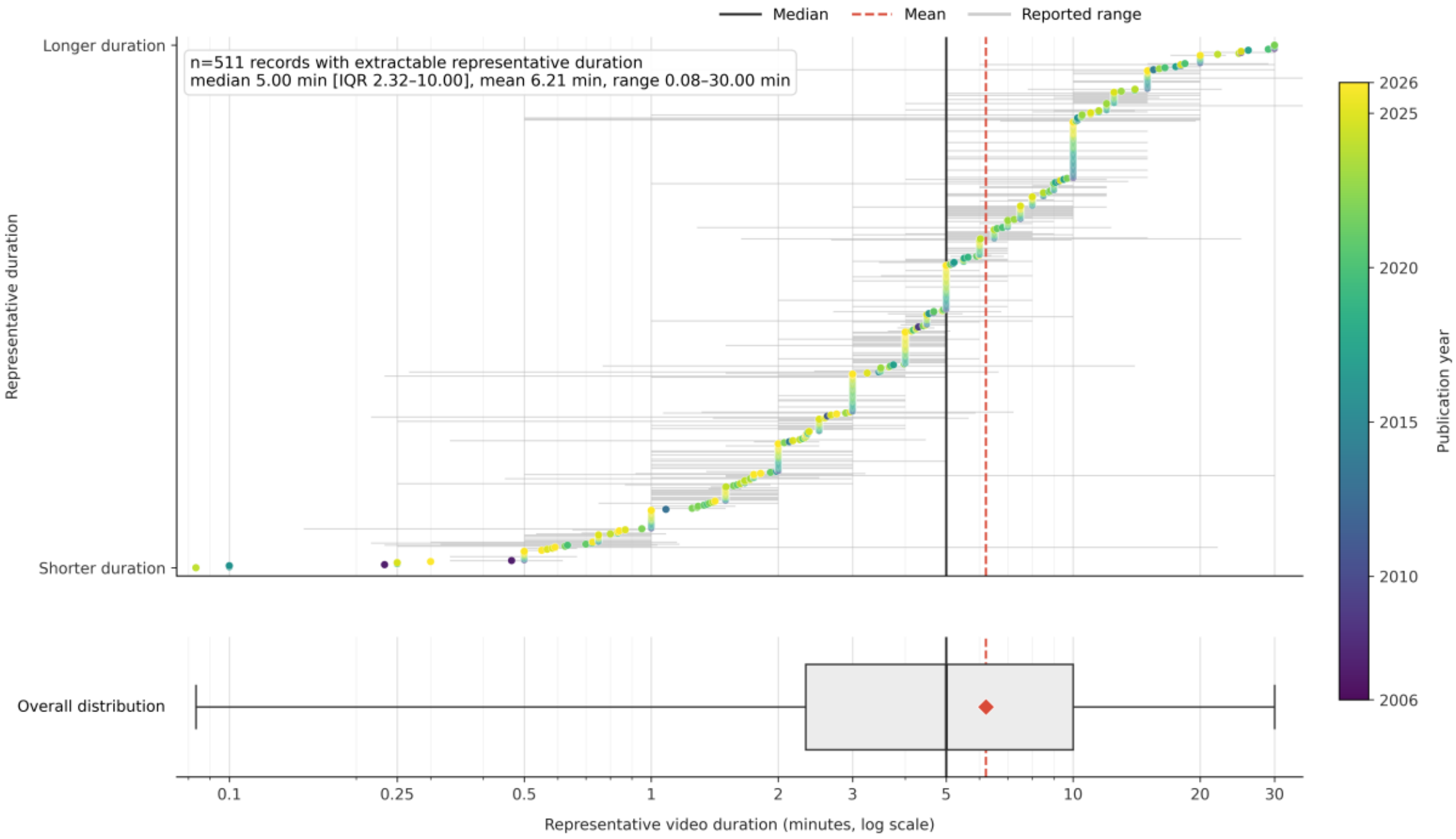

Figure 11. Distribution of extracted representative video durations.

Reported educational characteristics were examined across all 2,169 records using titles, abstracts and keywords. The resulting frequencies indicate which characteristics were foregrounded in indexed descriptions rather than providing exhaustive accounts of the underlying videos or learning activities.

The educationalRole dimension was identified most frequently, appearing in 1,528 records (70.5%). It was followed by educationalUse in 1,265 records (58.3%), teaches in 1,166 (53.8%), and interactivityType in 1,029 (47.4%). PlatformContext was identified in 729 records (33.6%), assesses in 541 (24.9%), educationalLevel in 339 (15.6%), and typicalAgeRange in 150 (6.9%) (Figure 12).

Within educationalRole, student was the most frequently identified category, followed by teacher, general public and professional. Within educationalUse, concept explanation was the most common category, followed by engagement or motivation, practice or reinforcement, assessment or feedback activity, and procedural demonstration. Within teaches, knowledge or content outcomes were identified most frequently, while affective or attitudinal outcomes and procedural or practical skill outcomes were reported less frequently.

The interactivityType dimension included both expositive and active categories, indicating that educational short videos were described both as presentation resources and as components of response, practice, discussion or production activities. Within platformContext, classroom or blended settings were identified most frequently, followed by course-management platforms, social short-video platforms and general video-sharing platforms. Within assesses, knowledge or comprehension outcomes were the most frequent, followed by reflective or metacognitive outcomes and procedural or skill-performance outcomes.

EducationalLevel and typicalAgeRange were less frequently foregrounded in indexed descriptions. These lower frequencies indicate that the information was less consistently available for corpus-wide comparison and should not be interpreted as evidence that the corresponding characteristics were absent from the underlying studies.

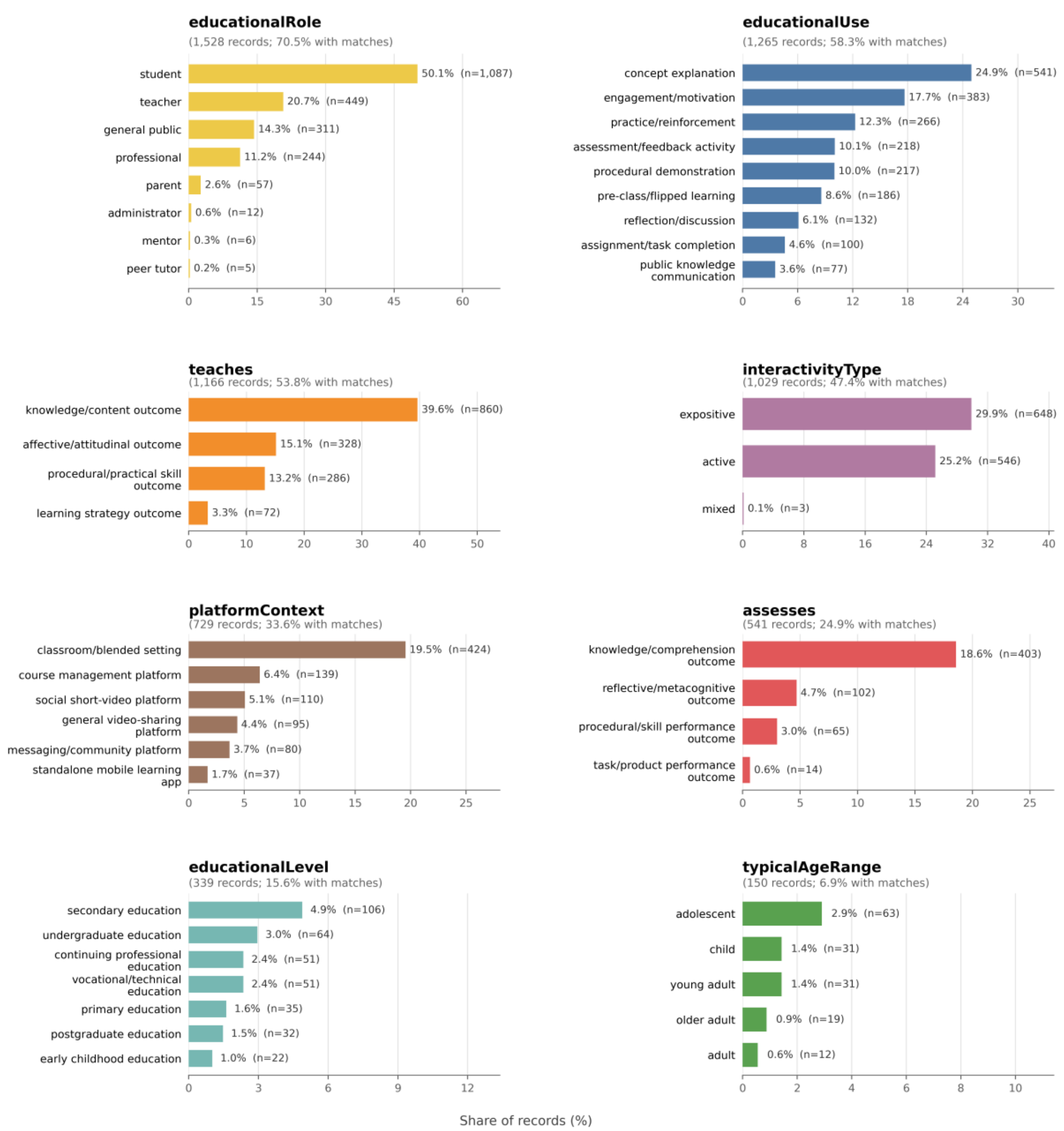

Figure 12. Reported characteristics of educational short videos in indexed record descriptions.

Taken together, the findings show that educational short video has not been operationalised through a stable duration boundary. Indexed descriptions also situated short videos within different educational configurations, including explanation, demonstration, practice, assessment and learner production. Studies using the same label may therefore refer to resources that differ in both temporal form and educational function. Meaningful comparison requires shortness to be considered alongside educational purpose, content unit, learner activity and context, rather than treated as a fixed duration category.

# 4. Discussion

This study examined the research trends, thematic organisation, and operationalisation of educational short-video research. The findings portray a rapidly growing but heterogeneous research area. Publications were widely dispersed across outlets, and many international collaboration links appeared in only one record. The 16 first-level topics differed in their size, connectivity, development, citation visibility, outcome coverage and reported study-design profiles. Reported video durations also varied substantially, while educational short videos were associated with different educational roles, uses, activities and contexts. Taken together, these patterns suggest that educational short video is better understood as a broad research domain than as a single and uniform educational intervention.

## 4.1 Growth across a fragmented field

Educational short-video research has developed across a wide range of disciplines and applications, including classroom instruction, language learning, professional training, health education and platform-based public education. This breadth is not inherently problematic. It reflects the adaptability of relatively brief video resources across educational settings. It may, however, make it more difficult to develop shared concepts and reporting practices across the field. The collaboration results point to a similar pattern, researchers from many countries contributed to the literature, but recurring international links were limited in the observed co-authorship network.

The thematic structure combined connection with differentiation. Skill Development in Educational Contexts had the highest node strength in the metadata-derived topic co-occurrence network, while other substantial areas focused on health interventions, patient education, instructional-video design, microlearning, and social-media engagement. The strongest overlaps generally occurred between neighbouring topics rather than across several application areas. The field was therefore not composed of isolated topics, but broader connections across distinct areas were less common than pairwise overlaps between closely related topics.

Topic size, network position, recent growth and citation visibility captured distinct dimensions of field development. Skill Development in Educational Contexts formed the broadest structural core, but its citation share remained below its publication share, showing that volume and connectivity do not necessarily translate into proportional citation visibility. Video-Based Health Interventions for Attitude Change and Cognitive Load and Engagement in Instructional Video Design combined recent growth with comparatively strong citation positions, whereas Social Media Engagement Strategies for Education expanded rapidly without a corresponding citation share, consistent with an emerging research direction and possible citation lag. These patterns indicate uneven development across topics and caution against judging topic maturity or significance from publication volume or growth alone.

## 4.2 Differences in reported study-design and outcome profiles

The topic–outcome analysis indicates that different areas of educational short-video research address different educational processes. Knowledge/Achievement and Engagement/Motivation were widely represented, while some topics placed greater emphasis on procedural performance, cognition, assessment or content quality. The identifiable study-design profiles also varied. Experimental and synthesis-level designs appeared more frequently in health-related topics, whereas descriptive and non-randomised designs were more common in several other areas.

These differences should not be interpreted as a hierarchy of topic quality. Study design could be identified for only 48.5% of the records, the classification relied on indexed metadata, and the study did not assess methodological quality or risk of bias. The findings instead suggest that different research designs are better suited to supporting different kinds of claims. Experimental studies can address comparative effects under specified conditions, while descriptive, qualitative, developmental and content-analytical studies can provide evidence about implementation, learner experience, emerging practice and resource characteristics.

Studies should therefore not be treated as comparable simply because they examine videos described as short. Measures of viewing or participation may indicate engagement, but they do not in themselves demonstrate understanding, retention or transfer (Guo et al., 2014). An instructor explanation, a learner-generated artefact and a procedural training video also involve different learning activities. The general question of whether short videos are effective is difficult to answer without first specifying the video's educational function, the activity expected of learners, the context of use and the outcome being examined.

## 4.3 Defining educational short video

The duration analysis shows that *short* has no stable numerical meaning across the literature. Among the 511 records for which duration could be extracted, reported values ranged from a few seconds to 30 minutes, with no consistently applied threshold for classification. Related research does not resolve this ambiguity. The six-minute recommendation proposed by Guo et al. (2014) was specific to learner engagement in a particular MOOC context, whereas microlearning studies have reported recommended durations ranging from 1–3 minutes to 10–15 minutes (Monib et al., 2025). Similarly, the 15-second-to-3-minute range reported in TikTok studies reflected the platform format examined at the time rather than an educational criterion (Paksoy et al., 2023). Because these values derive from different instructional purposes, research contexts, and platform conventions, they are better understood as context-specific recommendations or descriptive ranges than as general definitions of educational short video (Seidel, 2024).

The educational-characteristics analysis clarifies what duration alone leaves unspecified. Across the corpus, indexed descriptions associated educational short videos with different intended users, educational uses, learning content, forms of interactivity and platform contexts. Some were positioned as resources for explanation or demonstration, whereas others formed part of practice, assessment, discussion or learner-production activities. These distinctions concern the role of the video within an educational process, which cannot be inferred from its length. A fixed duration threshold would therefore place temporally similar but educationally different resources in the same category without distinguishing how they were intended to support learning. Duration is better treated as one attribute of an educational configuration than as the sole criterion defining the object.

Therefore, building on Mayer's (2020, 2021) concept of multimedia instructional messages, we define educational short videos as discrete multimedia messages that combine words and visuals and are designed or deliberately used to promote learning. Their shortness is relational rather than fixed, it is judged in relation to the educational purpose, content unit and surrounding activity in which the video is embedded, rather than by a universal duration threshold.

This definition establishes three minimum criteria: first, the video must constitute an identifiable and discrete audiovisual unit; second, it must perform an identifiable educational function; and third, its shortness must be interpretable in relation to its intended purpose and context of use. These criteria provide a common basis for identifying educational short videos without implying that all videos meeting the definition are pedagogically equivalent.

## 4.4 Implications for research and practice

Primary studies should report video duration or duration range alongside educational purpose, intended learners, expected learner activity, instructional or platform context, study design and outcomes. Duration should not be treated as an isolated intervention characteristic. Research should instead examine how its educational implications vary with content scope, segmentation, pacing, interactivity, learner control and surrounding learning activities. Further research should also determine whether duration and duration-reporting practices differ across educational functions, disciplines and platform contexts.

For systematic reviews and meta-analyses, studies should not be grouped solely because they use the label *short video* or report similar durations. Comparability should be assessed in relation to educational function, learner activity, context, study design and outcome measurement. Studies should be combined only when these characteristics are sufficiently aligned; otherwise, they should be examined as potential sources of heterogeneity.

For educators and instructional designers, the findings do not support a universal recommendation for video length. Duration decisions should reflect instructional purpose, content complexity and coherence, learners' prior knowledge and the activity in which the video is embedded. Descriptive duration statistics should not be interpreted as design rules. The practical priority is to scope and structure each video around a coherent learning unit and integrate it with what learners are expected to understand or do.

## 4.5 Limitations

This study was limited to records indexed in Web of Science and Scopus. Relevant publications indexed elsewhere, published in outlets with more limited database coverage, or described using terminology not captured by the search strategy may have been omitted. The corpus should therefore be regarded as a broad representation of indexed educational short-video research rather than an exhaustive account of the field. Citation counts were database-specific and indicate visibility within each database rather than complete scholarly influence.

Topic modelling and structured classifications relied primarily on titles, abstracts and keywords rather than full article texts. This provided a consistent representation of all 2,169 records but restricted the analysis to topics and characteristics foregrounded in indexed descriptions. Detailed procedures, secondary outcomes and learning activities reported only in the full text may not have been captured.

Study design could be identified for 48.5% of the corpus. Failure to identify a category from the available metadata should not be interpreted as evidence that the corresponding information was absent from the underlying study. Although the educational-characteristics dictionary underwent internal validation, this did not establish the completeness of metadata-based coding relative to full-text assessment. The study-design and outcome-related classifications were not independently validated against full-text coding. Their topic-level distributions may therefore reflect differences in reporting completeness as well as differences among the underlying studies and should be interpreted as exploratory metadata-based profiles rather than exhaustive classifications.

The NMF topic solution was influenced by preprocessing decisions, the selected number of topics and the threshold used to represent overlapping topic membership. Topic labels also required researcher interpretation and should be understood as concise descriptions of model-derived components rather than fixed or mutually exclusive disciplinary categories.

A representative duration could be extracted for 511 records, corresponding to 23.6% of the corpus. The duration findings therefore describe only records containing sufficiently explicit and quantifiable information. They cannot be assumed to represent the duration distribution of records for which such information was unavailable.

Finally, this bibliometric, topic-modelling and structured content analysis did not estimate pooled intervention effects or conduct study-level risk-of-bias assessments. The findings describe the structure, reporting patterns and operationalisation of the literature and should not be interpreted as evidence of educational effectiveness.

## 5. Conclusion

Educational short-video research constitutes a broad and internally differentiated domain rather than a single intervention category. The 16 first-level topics were connected but showed distinct patterns of development, citation visibility, study design and outcome coverage. Across this literature, short videos described using the same terminology also differed in their educational purposes, intended users, learner activities and contexts. They should therefore not be assumed to represent educationally equivalent resources or directly comparable interventions.

The findings further show that short has not been operationalised through a consistently applied numerical boundary. We therefore propose a working definition of educational short videos as discrete multimedia messages that combine words and visuals and are designed or deliberately used to promote learning. Their shortness is relational and should be interpreted in relation to educational purpose, content unit and surrounding activity rather than determined by a universal duration threshold. This definition provides a common basis for identifying the object of study while retaining meaningful distinctions in how videos are designed and used. Future primary research and evidence synthesis should assess comparability through these educational configurations, while educators should select and structure videos according to what learners are expected to understand or do.